\documentclass[3p,review,sort&compress]{elsarticle}
\usepackage{amsmath}
\usepackage{amssymb}
\usepackage{amsfonts}
\usepackage{epsfig}
\usepackage{graphicx}
\usepackage{multirow}
\usepackage{caption}
\usepackage{cite}
\usepackage{float}

\journal{ArXiv}

\begin{document}

\begin{frontmatter}

\title{Cooperation in the two-population snowdrift game\\ with punishment enforced through different mechanisms}

\author[PUC]{Andr\'e Barreira da Silva Rocha\corref{cor1}}
\ead{andre-rocha@puc-rio.br}
\cortext[cor1]{corresponding author}
\address[PUC]{Department of Industrial Engineering, Pontifical Catholic University of Rio de Janeiro,\\ Rua Marqu\^es de S\~ao Vicente, 225, G\'avea, CEP22451-900, Rio de Janeiro, RJ, Brazil.}

\begin{abstract}
I study two mechanisms based on punishment to promote cooperation in the two-population snowdrift game. The first mechanism follows the traditional approach in the literature and is based on the inclusion of a third additional strategy in the payoff matrix of the stage-game. The second mechanism consists of letting cooperators to punish defectors with a given exogenous frequency. While both mechanisms share the same result regarding the minimum required level of punishment in order to eliminate defectors in both populations, stability in the mechanism following the second approach is more robust in the sense that extinction of defectors is a globally asymptotically stable state for any interior initial conditions in the phase space. Results were obtained analytically through non-linear differential equations and also using an agent-based simulation. There was a good level of agreement between both approaches with respect to the evolutionary pattern over time and the possible steady-states.    
\end{abstract}


\end{frontmatter}

\section{Introduction}
\label{sec:intro}
The emergence of cooperation in games played by individuals drawn from the same population has been widely studied in the literature across many branches of science \citep{Epstein96, Riolo01, Hauert02, Hauert04, Hauert05, Nowak06, Zhong06, Xu11, Wardil10, Requejo13, Chan13}. Typically, focus relies on the Prisoner's Dilemma (PD) and on the Snowdrift Game (SG). In well-mixed populations, defectors take over in the former class of game while defectors and cooperators co-exist in the latter class, although in the SG there is still a paradox due to the fact that the average fitness of the population is lower than it would be if only cooperators survived in the steady-state. Thus, mechanisms to increase the frequency of cooperation have been studied for both classes of games. 

One popular mechanism is the inclusion of spatial structures such as spatial lattices with periodic boundary conditions and Moore neighbourhood where each individual interacts only with the eight nearest neighbours, reachable by a chess-king's move \citep{Hauert04}. While in the PD structured populations tend to increase cooperation, in the SG defectors are favoured \citep{Hauert04, Hauert05}. A second popular mechanism is to keep the population well-mixed and include additional strategies in the payoff matrix of the stage-game. Examples of additional strategies in the SG are the Bourgeois \citep{Maynard82}, the Retaliator \citep{Zeeman81,Bomze83} and the Punisher \citep{Xu11}. In the latter, an individual programmed to play the Punisher strategy acts as a cooperator when facing an opponent willing to cooperate but acts as punisher when facing a defector, i.e., the individual pays a cost to punish such that the defector also incurs a loss which is larger than the cost. Punishers transform cooperators into a type of free-rider where the former pay the costs while the latter enjoy the benefits of the harm done to defectors.

In this paper, I use the framework of evolutionary game theory (EGT) to analytically study the SG. Differently from the usual literature, I study the SG played by individuals drawn from two different well-mixed populations. In this case, although in the steady-state cooperation and defection co-exist as in the SG played by a single population, there are two steady-states instead of a single one. In both steady-states, one population ends up only with cooperators and the opponent population contains only individuals programmed to defect. I then study two mechanisms based on punishment in order to enforce full cooperation in both populations. The first mechanism is an extension of \citep{Xu11}, i.e., I include a third strategy in the strategy set of both populations. The dynamic model can be studied in the $\Re^4$-space and results are in line with \citep{Xu11} in the sense that extinction of defectors in both populations may be a stable steady-state in which only neutral stability is attained and the existence of more than one basin of attraction implies that the elimination of defectors depends on the initial conditions. In the second mechanism, instead of including additional strategies into the stage-game, the pure strategy Cooperation is replaced by a new mixed strategy which I name Cooperation-Punishment (CP). The essence of this new strategy is that with a given exogenous probability, a cooperator might punish a defector. Depending on the frequency of punishment, the penalty on defectors and the cost to punish, the SG may have the best response structure of its payoff matrix transformed for example into a class of coordination game or a game with dominant strategy other than a PD. While in the latter case, defectors become worse off, in the former extinction of defectors is feasible depending on the initial conditions. 

The advantages of the second mechanism of punishment are twofold: first, from a mathematical point of view, the dynamics is kept in the more intuitive $\Re^2$-space as in \citep{Xu11}'s 3-strategy SG played by one single well-mixed population; second, from an evolutionary point of view, differently from the mechanism that relies on the introduction of a new character who always punishes defectors, extinction of defectors is independent from the initial conditions and, differently from \citep{Xu11}, stability is asymptotic. Neutral invasion of any population is not possible anymore. Despite the differences pointed out above, both mechanisms still share one important aspect of \citep{Xu11}, which is the minimum required level of punishment in order to eliminate defectors in a population. The latter is satisfied as long as the frequency of punishment is above $(r/\beta)$, where $r$ is the cost-to-net-benefit ratio as defined in \citep{Hauert05} and $\beta$ is the penalty a defector faces when punished.

In other words, the second mechanism displays a simpler model where cooperation, punishment and defection are possible behaviours, still keeps the same threshold as in \citep{Xu11} to eliminate defection, but the robustness of the equilibrium in which defection becomes extinct is improved.

The deterministic results obtained through EGT are then confirmed in a stochastic agent-based simulation where I extend the strategy updating criterion used in \citep{Zhong06,Xu11,Wardil10} to the case of agents selected from two different populations. To my best knowledge, this is the first time in the literature that such simulation is carried out with two competing populations. A target agent obtaining a lower payoff than the referencing agent may update his strategy probabilistically. During each Monte Carlo time step, on average, all agents have the chance to update their strategy. Strategies are updated asynchronously. I show that both the analytical model and the simulations display good agreement regarding the steady-states and the evolutionary patterns.

The rest of the paper is organized as follows. In Section 2, I review the SG played by one population and introduce the SG played by two populations as well as the two different mechanisms to enforce cooperation based on punishment. The analysis uses EGT based on the replicator dynamics (RD). In Section 3 the agent-based simulation is carried out and results are compared with the analytical results of the previous section. Section 4 concludes. 
 
\section{Snowdrift game: one population and two populations}
\label{sec:SG}
I depart from a SG repeatedly played over time between two randomly selected players, each belonging to one of two very large different populations, the latter named populations $1$ and $2$. Thus, individuals drawn from one population only interact with opponents drawn from the other population, i.e., own-population effects are not taken into account in the two-population game. In the classic two-player SG, each player has two pure available strategies, either to cooperate ($C$) or to defect ($D$). When both players cooperate, each receive a reward payoff $R=b-(c/2)$ resulting from the benefit of cooperation $b$ net of the cost to cooperate $c$, which is shared by both players. On the other hand, when only one player cooperates, the cooperator bears the entire cost $c$, getting a ``sucker'' payoff $S=b-c$, while the opponent earns the costless temptation payoff $T=b$. If neither player cooperates, both earn nothing, i.e., they get each $P=0$. Given $b>c>0$ in the SG, we have $T>R>S>P$. This can be represented by the following payoff matrix:
\begin{equation}
\begin{bordermatrix} {&C&D\cr
		C & b-c/2,b-c/2 &b-c,b\cr
		D & b,b-c&0,0\cr} 
		\end{bordermatrix}
\label{Matrix-I}
\end{equation}
Multiplying all entries in the previous matrix by $\frac{2}{2b-c}$, it can be transformed into an equivalent matrix in which payoffs depend on the cost-to-net-benefit ratio $r=\frac{c}{2b-c}$, where $0<r<1$:
\begin{equation}
\begin{bordermatrix} {&C&D\cr
		C & 1,1 &1-r,1+r\cr
		D & 1+r,1-r&0,0\cr} 
		\end{bordermatrix}
\label{Matrix-II}
\end{equation}
In well-mixed populations, based on the RD and on (\ref{Matrix-I}), the classic SG assumes that all individuals belong to the same population and the following ordinary differential equation (ODE) governs the evolution of the proportion of cooperators $x$ over time:
\begin{equation*}
\dot{x}=x(1-x)\left[(b-c/2)(1-x)-c/2 \right] 
\end{equation*}
leading to the set of stationary points $x=\left\lbrace 0,1-r,1 \right\rbrace $, where $x\in\Re\therefore\dot{x}=0$, from which only $\bar{x}=1-r$ is asymptotically stable, i.e., $\bar{x}\in\Re\therefore\frac{\partial \dot{x}}{\partial x}(\bar{x})<0$, thus an evolutionary equilibrium (EE) as defined in \citep{Friedman98}. As a consequence, in the SG played by individuals drawn from a well-mixed homogeneous population, cooperation emerges with a frequency $1-r$ and defection co-exists with the latter and has frequency $r$. For the SG played by individuals drawn from a well-mixed heterogeneous population, see \citep{Barreira13}. In \citep{Xu11}, the inclusion of a third pure strategy named punishment leads to the extinction of defectors if $r$ is below a given threshold, but generally the population evolves to a mixture of cooperators and punishers and stability of the equilibrium is not robust in the sense that only neutral stability, but not asymptotic stability, holds.

Keeping the stage-game as in (\ref{Matrix-I}) but assuming that each player comes from a different population, results change substantially. In such a case, the evolution of cooperation in each population is governed by the following non-linear system of ODEs:
\begin{eqnarray}
\dot{x_1}&=&x_1(1-x_1)\left[(b-c/2)(1-x_2)-c/2 \right]\label{RD-Ix}\\
\dot{x_2}&=&x_2(1-x_2)\left[(b-c/2)(1-x_1)-c/2 \right]\label{RD-Iy}
\end{eqnarray}
The phase space becomes the unit square $[0,1]\times[0,1]\in\Re^2$ and the stationary points are the corners of the phase space $(0,0)$, $(0,1)$, $(1,0)$, $(1,1)$ as well as there is an interior stationary point $(1-r,1-r)$. Analysing the linearised system in the neighbourhood of the stationary points, the eigenvalues of the Jacobian matrix evaluated at the corners of the phase space are $\lambda_1=\frac{\partial\dot{x_1}}{\partial x_1}$ and $\lambda_2=\frac{\partial\dot{x_2}}{\partial x_2}$, which are both negative at $(0,1)$ and $(1,0)$ and both positive at $(1,1)$ and $(0,0)$. On the other hand, the eigenvalues of the Jacobian matrix evaluated at $(1-r,1-r)$ are $\lambda_{1,2}=\pm r(1-r)$. Thus, the latter is a saddle point, $(1,1)$ and $(0,0)$ are sources and $(1,0)$ and $(0,1)$ are sinks. There are two basins of attraction and, differently from the SG played by one single population of agents, the system evolves to one of two asymptotically stable nodes leading to one population of only cooperators and one population of only defectors.

On the other hand, similar to the SG played by a single well-mixed population, defection always co-exists with cooperation when the SG is played by two populations, although only one strategy survives in each population. In the next two sub-sections, I look into two mechanisms to enforce extinction of defection in both populations. The first mechanism follows the approach in \citep{Xu11} and consists in the introduction of an additional pure strategy in both populations in order to punish defectors with certainty. Although it is possible to eliminate defectors completely, this evolutionary pattern depends on the initial conditions and on the parameters $\beta$ and $r$. I then look into a second mechanism, consisting of a change of behaviour of the individuals programmed to cooperate. I assume that cooperators may also punish defectors with a certain exogenous probability. The latter mechanism is equivalent to replace the pure strategy Cooperation by a mixed strategy where the individual randomizes his choice between Cooperation and Punishment. It has the mathematical advantage that the dynamics is kept in the more intuitive $\Re^2$-space which, as in the system defined by (\ref{RD-Ix}) and (\ref{RD-Iy}), makes it possible to keep the use of a planar phase space to study the evolutionary pattern of both populations as well as the possible dependence of each pattern on the initial conditions. Moreover, as will be shown in this Section, differently from the mechanism that relies on the introduction of a third character who always punishes defectors as in \citep{Xu11}, stability is asymptotic whenever defectors become extinct in both populations. 

\subsection{Punishment enforced by a third character}
\label{sec:SG-I}
I start incorporating punishment into the SG by introducing a third pure strategy named punishment (P), in the same way it is done in \citep{Xu11} except that I deal with a two-population SG. In this setup, whenever a cooperator and a punisher meet it is the same as if two cooperators were meeting, thus both get the payoff $R=1$ as in (\ref{Matrix-II}). On the other hand, when a punisher meets a defector, the former pays a cost $\alpha$ in order to enforce a penalty $\beta$ on the latter. Considering only the case $\beta>\alpha>0$, bounded rationality is still present in such a setup, i.e., the loss incurred by defectors is larger than the cost payed by punishers whenever the latter decide to punish. The payoff matrix in (\ref{Matrix-II}) then becomes:
\begin{equation}
\begin{bordermatrix} {&C&D&P\cr
		C & 1,1 &1-r,1+r&1,1\cr
		D & 1+r,1-r&0,0&1+r-\beta,1-r-\alpha&\cr
		P & 1,1&1-r-\alpha,1+r-\beta&1,1\cr} 
		\end{bordermatrix}
\label{Matrix-IV}
\end{equation}
The non-linear system representing the RD that governs the evolution of the shares of cooperators and defectors in populations $i,j=\left\lbrace 1,2\right\rbrace;\ i\neq j$ is:
\begin{eqnarray}
\dot{x_i}&=&x_i\left[ y_i(y_j-r)+\beta y_i(1-x_j-y_j)+\alpha y_j(1-x_i-y_i)\right]\label{RD-IIIx}\\
\dot{y_i}&=&y_i\left[ (1-y_i)(r-y_j)-\beta(1- y_i)(1-x_j-y_j)+\alpha y_j(1-x_i-y_i)\right]\label{RD-IIIy}
\label{RD-III}
\end{eqnarray}
where $x_i$ and $y_i$ are respectively the shares of cooperators and defectors in population $i$. The share of punishers present in the population is given by $1-x_i-y_i$. In the same vein of the system given by (\ref{RD-Ix}) and (\ref{RD-Iy}), there are two asymptotically stable states $(x_1,y_1,x_2,y_2)=(1,0,0,1)$ and $(0,1,1,0)$ where defectors take over one population and cooperators take over the opponent population. Also, similar to \citep{Xu11}, there is an invariant manifold $(x_1,0,x_2,0)$ containing a set of neutrally stable points that attract the vector field if $x_i<1-r/\beta;\ i=\left\lbrace 1,2\right\rbrace $\footnote{In contrast with \citep{Xu11}, here there is a planar invariant manifold while in the former the invariant manifold is linear.}. Any region of the invariant manifold that violates the previous condition repeals the vector field. Therefore, depending on the initial states of both populations, the system might evolve to a stable state with a mixture containing a share $x_i$ of cooperators and a share $1-x_i$ of punishers, i.e., defectors become extinct in both populations as long as the frequency of punishment is at least $1-x_i>(r/\beta);\ i=\left\lbrace 1,2\right\rbrace $. Note however that neutral invasion can still occur, i.e., if a small share of defectors invade the system, defection becomes extinct again but the shares of cooperators and punishers at the steady-state generally change.

The result can be confirmed from the Jacobian matrix $\Omega$ evaluated at those stationary points, where $i,j=\left\lbrace 1,2 \right\rbrace;\ i\neq j$ :
\begin{eqnarray*}
\frac{\partial \dot{x}_i}{\partial x_i}&=&y_i(y_j-r)+\beta y_i(1-x_j-y_j)+\alpha y_j(1-2x_i-y_i)\\
\frac{\partial \dot{x}_i}{\partial y_i}&=&x_i(y_j-r)+\beta x_i(1-x_j-y_j)-\alpha y_jx_i\\
\frac{\partial \dot{x}_i}{\partial x_j}&=&-\beta x_iy_i\\
\frac{\partial \dot{x}_i}{\partial y_j}&=&x_iy_i(1-\beta)+\alpha x_i(1-x_i-y_i)\\
\frac{\partial \dot{y}_i}{\partial x_i}&=&-\alpha y_iy_j\\
\frac{\partial \dot{y}_i}{\partial y_i}&=&(1-2y_i)(r-y_j)-\beta(1-2 y_i)(1-x_j-y_j)+\alpha y_j(1-x_i-2y_i)\\
\frac{\partial \dot{y}_i}{\partial x_j}&=&\beta y_i(1-y_i)\\
\frac{\partial \dot{y}_i}{\partial y_j}&=&-y_i(1-y_i)(1-\beta)+\alpha y_i(1-x_i-y_i)
\end{eqnarray*}

Based on the previous set of equations, the Jacobian matrix evaluated at $(1,0,0,1)$ leads to the set of eigenvalues $\lambda^{(1,0,0,1)}=\left\lbrace -\alpha,-(1-r),-r,-r\right\rbrace $, thus all eigenvalues are negative and $(1,0,0,1)$ is always asymptotically stable. The stationary state $(0,1,1,0)$ is also asymptotically stable given $\Omega(0,1,1,0)$ leads to the same set of negative eigenvalues.

\begin{figure}[htp]
\centering
\begin{tabular}{cc}
\epsfig{file=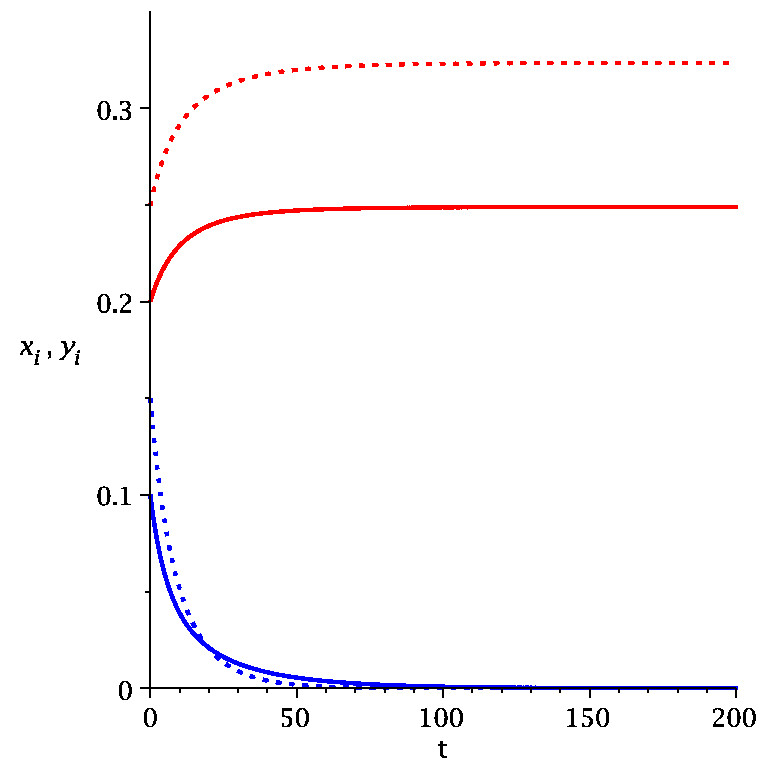,height=5cm,width=5cm,angle=0}&
\epsfig{file=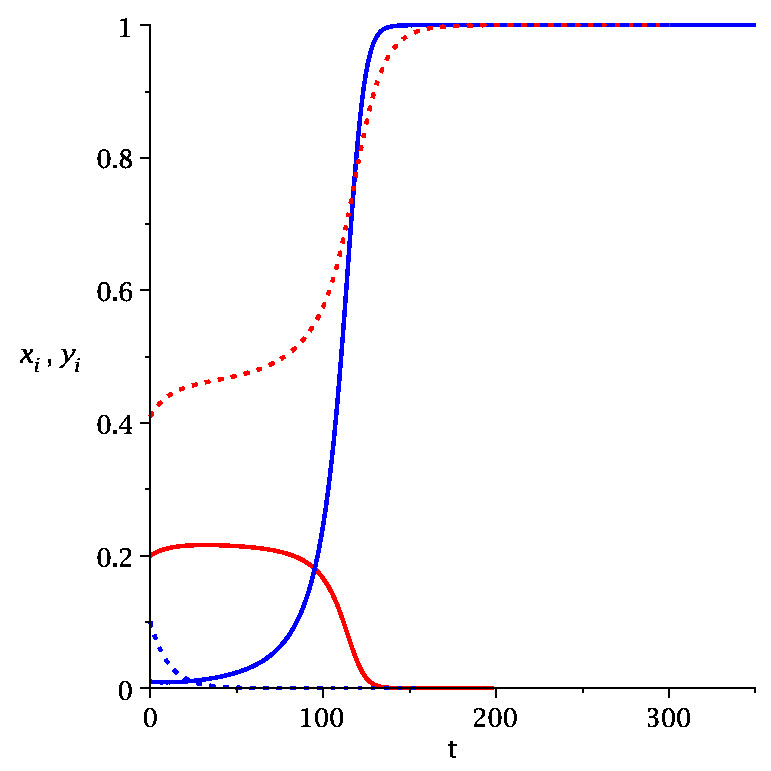,height=5cm,width=5cm,angle=0}
\end{tabular}
\vspace{.5cm}
\caption{(color online) left: for initial conditions $(0.20,0.10,0.25,0.15)$, $\alpha=0.1$, $\beta=0.5$ and $r=0.3$, cooperators (red) and punishers take over both populations while defectors (blue) become extinct. Right: for initial conditions $(0.2,0.01,0.41,0.10)$, $\alpha=0.1$, $\beta=0.5$ and $r=0.3$, defectors take over population 1 (solid line) while cooperators take over population 2 (dot line).}
\label{fig:1}
\end{figure}

For the invariant manifold $(x_1,0,x_2,0)$, the Jacobian matrix gives the set of eigenvalues $\lambda^{(x_1,0,x_2,0)}=\left\lbrace 0,r-\beta(1-x_1),0,r-\beta(1-x_2)\right\rbrace $. Stability requires non-zero eigenvalues to be negative, which is satisfied if $x_i<1-\frac{r}{\beta};\ i=\left\lbrace 1,2\right\rbrace $. Although the system is non-hyperbolic at nodes $(x_1,0,x_2,0)$ and the condition obtained from the Jacobian matrix is not sufficient to guarantee stability, this can be confirmed from the numerical simulations in Figure \ref{fig:1}. In both panels of the figure, the initial state of each population contains a mixture of cooperators, defectors and punishers where defectors are relatively rare, i.e., the initial condition is located sufficiently close to the invariant manifold $(x_1,0,x_2,0)$. In the left panel, the condition for the extinction of defectors $(x_1<0.4\wedge x_2<0.4)$ is satisfied. In the right panel, this condition is violated and, even with an initial state in which defectors are a residual share of just 1\% of population 1, the latter take over that population.

\subsection{Punishment enforced by a change of behaviour of cooperators}
\label{sec:SG-II}
In this subsection, I incorporate punishment into the SG by changing the way a cooperator individual behaves. I assume the pure strategy to cooperate $C$ is replaced by a mixed strategy named cooperation-punishment ($CP$) according to which an individual always behaves as a cooperator when meeting another individual programmed to play $CP$ but, when facing a defector, there is a probability $p_i;\ i=\left\lbrace 1,2 \right\rbrace$ that the cooperator will punish the defector by paying a cost to punish $\alpha$ and leading defectors to incur a loss $\beta$. In line with Subsection \ref{sec:SG-I}, $\beta>\alpha>0$. Under the perspective of the row and column players being respectively drawn from populations $1$ and $2$, after punishment is incorporated into the model, (\ref{Matrix-II}) becomes:  
\begin{equation}
\begin{bordermatrix} {&CP&D\cr
		CP & 1,1 &1-r-p_1\alpha,1+r-p_1\beta\cr
		D & 1+r-p_2\beta,1-r-p_2\alpha&0,0\cr} 
		\end{bordermatrix}
\label{Matrix-III}
\end{equation}
Note that by incorporating punishment into the cooperators' way of behaving, the dimensionality of the non-linear system governing evolution is kept in the $\Re^2$ as in (\ref{RD-Ix}) and (\ref{RD-Iy}). Without loss of generality, I assume that cooperator individuals belonging to population $1$ tend to punish defectors more often than cooperators from population $2$, such that $p_1=p_2+\delta,\ \delta>0$.  

From (\ref{Matrix-III}), after punishment is incorporated, (\ref{RD-Ix}) and (\ref{RD-Iy}) become respectively:
\begin{eqnarray}
\dot{x_1}&=&x_1(1-x_1)\left[1-r-x_2-p_1\alpha(1-x_2)+x_2p_2\beta  \right]\label{RD-II-x}\\
\dot{x_2}&=&x_2(1-x_2)\left[1-r-x_1-p_2\alpha(1-x_1)+x_1p_1\beta  \right]\label{RD-II-y}
\end{eqnarray}
leading to four stationary points at the corners of the phase space and an interior stationary point $(\overline{x}_1,\overline{x}_2)=\left(\frac{1-p_2\alpha-r}{1-p_2\alpha-p_1\beta},\frac{1-p_1\alpha-r}{1-p_1\alpha-p_2\beta} \right)$, which might or not belong to $[0,1]^ 2$. Punishment introduced as in (\ref{RD-II-x}) and (\ref{RD-II-y}) allows for very rich dynamics, including two different cases with an interior saddle node. The eigenvalues of the Jacobian matrix $\Omega$ evaluated at $(0,0)$, $(1,1)$, $(1,0)$ and $(0,1)$ are respectively:
\begin{eqnarray}
\lambda_1^{(0,0)}=1-r-p_1\alpha& &\lambda_2^{(0,0)}=1-r-p_2\alpha\label{eig1}\\
\lambda_1^{(1,1)}=r-p_2\beta& &\lambda_2^{(1,1)}=r-p_1\beta\label{eig2}\\
\lambda_1^{(0,1)}=-r+p_2\beta& &\lambda_2^{(0,1)}=-(1-r)+p_2\alpha\label{eig3}\\
\lambda_1^{(1,0)}=-(1-r)+p_1\alpha& &\lambda_2^{(1,0)}=-r+p_1\beta\label{eig4}
\end{eqnarray}  

The eigenvalues of $\Omega$ evaluated at the interior stationary state $(\overline{x}_1,\overline{x}_2)$ are given by are $\lambda_{1,2}^{(\overline{x}_1,\overline{x}_2)}=\pm\left[ \overline{x}_1\overline{x}_2(1-\overline{x}_1)(1-\overline{x}_2)(1-p_1\alpha-p_2\beta)(1-p_2\alpha-p_1\beta)\right] ^{1/2}\in\Re$. For pure imaginary eigenvalues, $1-p_1\alpha-p_2\beta>0\wedge 1-p_2\alpha-p_1\beta<0$, in which case $(\overline{x}_1,\overline{x}_2)\in[0,1]^2$ requires $p_2\alpha>1-r>p_1\alpha \wedge p_1\beta>r>p_2\beta$, which cannot hold simultaneously. Thus, there is no possibility of cyclic behaviour with closed orbits about the interior neutrally stable state, in which case an evolutionary equilibrium would not exist.

\textbf{Definition 1:} given the cost to punish $\alpha$, the penalty on defectors $\beta$, the cost-to-net-benefit ratio $r$ and the probability $p_i,\ i=\left\lbrace 1,2\right\rbrace $ that a cooperator from population $i$ enforces punishment on a defector from the opponent population, the following types of punishment can be found in population $i$:
\begin{itemize}
\item $p_i\beta>r\wedge(1-r)<p_i\alpha$ (harsh punishment): the likelihood of punishment is such that, on average, the extra reward $r$ a defector gets when facing a cooperator is more than offset by the expected received punishment $p_i\beta$. But on the other hand, the likelihood of punishment also penalises cooperators given that the expected cost to punish leads to an expected negative payoff when the latter meet a defector, i.e., $(1-r)-p_i\alpha<0$;
\item $p_i\beta<r\wedge(1-r)>p_i\alpha$ (soft punishment): the opposite from harsh punishment, cooperators are very lenient such that the likelihood of punishment is low enough to grant both defectors and cooperators with an expected positive payoff when they meet each other;
\item $p_i\beta<r\wedge(1-r)<p_i\alpha$ (inefficient punishment): the likelihood of punishment is such that, on average, defectors end up with a positive payoff when they meet a cooperator but the latter gets a negative payoff;
\item $p_i\beta>r\wedge(1-r)>p_i\alpha$ (efficient punishment): on average, cooperators are able to enforce a negative payoff on defectors when both meet but the former still get a positive payoff even after incurring the expected cost to punish.
\end{itemize}

\textbf{Theorem 1:} there are five feasible cases regarding the evolutionary equilibria: (i) in both populations, cooperators take over while defectors become extinct if either both populations enforce efficient punishment or efficient punishment is enforced by population 2 and harsh punishment is enforced by population 1; (ii) in both populations, defectors take over if either both populations enforce inefficient punishment or population 2 enforces inefficient punishment and population 1 enforces harsh punishment; (iii) an equilibrium with only cooperators in both populations co-exists with another equilibrium with only defectors in both populations whenever both populations enforce harsh punishment; (iv) two equilibria where cooperators take over one population and defectors take over the other population co-exist whenever both populations enforce soft punishment. This case is similar to the standard SG played by two populations without punishment; (v) an equilibrium with cooperators taking over one population and defectors taking over the other population holds whenever population 2 enforces soft punishment and population 1 enforces any type of punishment other than the soft one.

\textbf{Proof:} from (\ref{eig1}-\ref{eig4}), in case (i), if both populations enforce efficient punishment (in this case the SG becomes a game in which strategy $CP$ is dominant), $(0,0)$ is a source, $(1,1)$ is a sink and both $(0,1)$ and $(1,0)$ are saddle points. If population 1 enforces harsh punishment instead, $(0,0)$ and $(0,1)$ are both saddle points, $(1,1)$ is a sink and the corner $(1,0)$ is a source; (ii) if both populations enforce inefficient punishment, the SG becomes a PD, i.e., $D$ is a dominant strategy and all eigenvalues in (\ref{eig1}-\ref{eig4}) change sign, thus $(0,0)$ and $(1,1)$ become a sink and a source respectively. If population 1 enforces harsh punishment instead, $(1,1)$ becomes a saddle point. The corners $(1,0)$ and $(0,1)$ behave as in case (i); (iii) when both populations enforce harsh punishment, the SG becomes a coordination game and $(0,0)$ and $(1,1)$ are sinks, $(1,0)$ and $(0,1)$ are sources and $(\bar{x}_1,\bar{x}_2)$ is an interior saddle point given that $1-r<p_i\alpha;\ i=\lbrace 1,2\rbrace$, which requires $1-p_i\alpha-p_j\beta<0;\ i,j=\lbrace 1,2\rbrace, i\neq j$ such that $(\bar{x}_1,\bar{x}_2)\in\left[0,1 \right]^2 $, leading to a Jacobian matrix $\Omega$ with real eigenvalues of opposite sign; (iv) if both populations enforce soft punishment, all eigenvalues in (\ref{eig1}-\ref{eig4}) change sign relatively to case (iii), thus $(1,0)$ and $(0,1)$ become sinks while $(1,1)$ and $(0,0)$ are sources; (v) when population 2 enforces soft punishment and population 1 enforces a punishment other than the soft one, $(0,0)$, $(1,1)$ and $(1,0)$ are respectively a source, saddle point and saddle point (efficient punishment), a saddle point, source and saddle point (inefficient punishment), a saddle point, saddle point and a source (harsh punishment), while the corner $(0,1)$ is always a sink. 
$\bullet$   

\begin{figure}[htp]
\centering
\begin{tabular}{cccc}
\epsfig{file=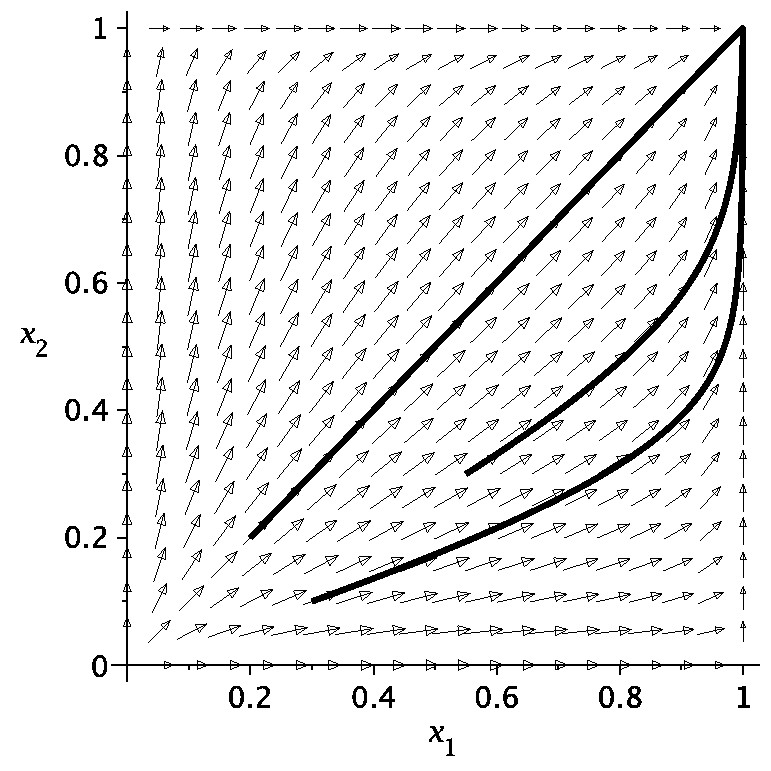,height=4cm,width=3.7cm,angle=0}&
\epsfig{file=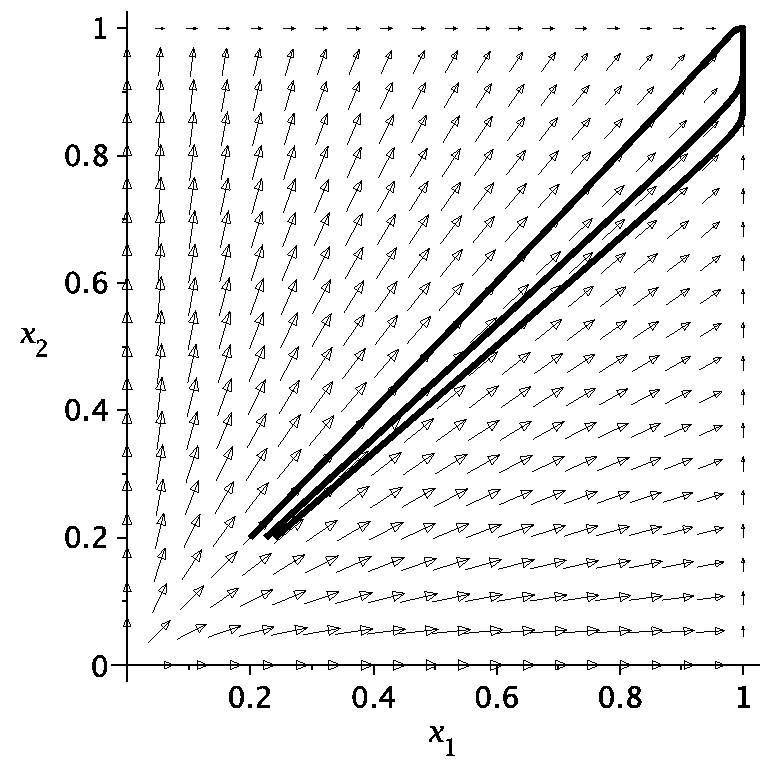,height=4cm,width=3.7cm,angle=0}&
\epsfig{file=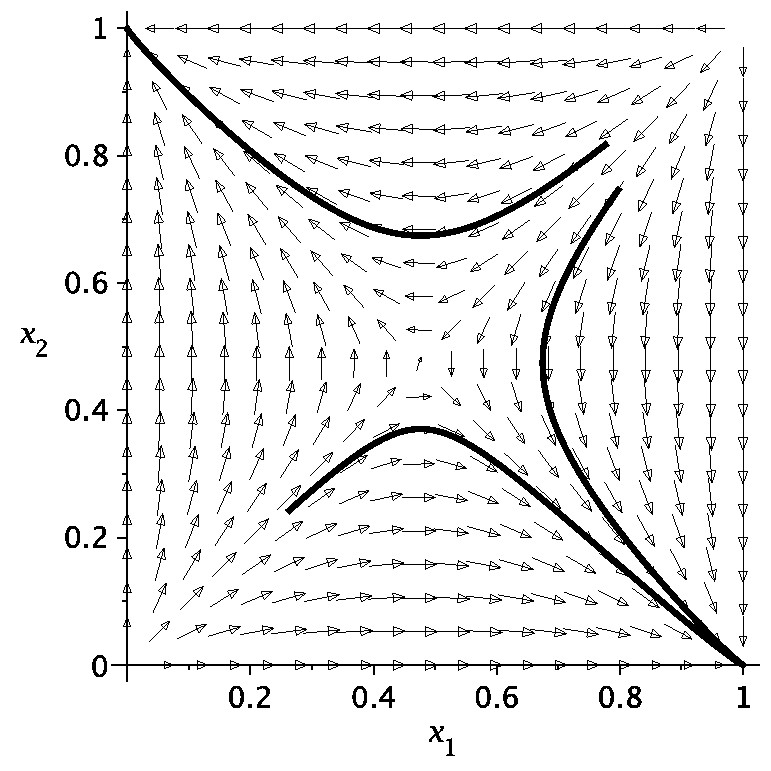,height=4cm,width=3.7cm,angle=0}&
\epsfig{file=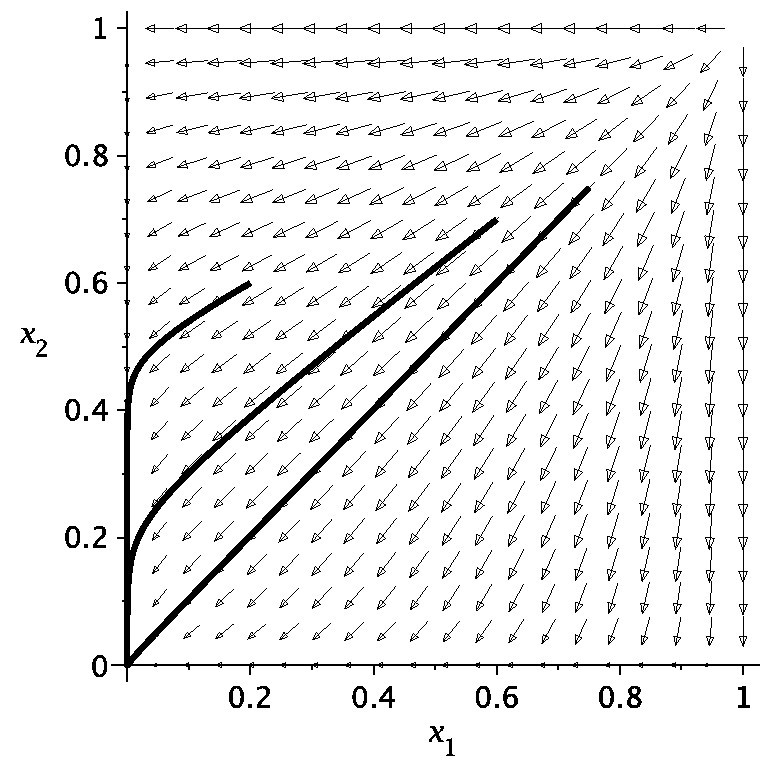,height=4cm,width=3.7cm,angle=0}
\end{tabular}
\vspace{0.5cm}
\caption{from left to right, phase space for $\alpha=0.1$, $\beta=0.5$, $p_1=0.41$, $p_2=0.405$ and respectively $r=0.05$ (both pops. enforce efficient punishment), $r=0.20$ (both pops. enforce efficient punishment), $r=0.60$ (both pops. enforce soft punishment) and $r=0.99$ (both pops. enforce inefficient punishment).}
\label{fig:2}
\end{figure}

In Figure \ref{fig:2}, I present a simulation using $\alpha=0.1$, $\beta=0.5$, both parameters as in \citep{Xu11}, $p_1=0.41$, $p_2=0.405$ and $r$ varying from $0.05$ to $0.99$. In \citep{Xu11}, for $r$ below a given threshold ($\bar{r}\approx 0.20$), defectors become extinct and a mixture of cooperators and punishers survive. When $r=\bar{r}$, their numerical simulations indicate that in the steady-state the shares of cooperators and punishers are respectively $\lesssim 0.60$ and $\gtrsim 0.40$. Above the threshold, punishers are driven out of the population and cooperators and defectors co-exist keeping the same shares in the population as in the classic SG, i.e., respectively $1-r$ and $r$. Here, for $0<r\lesssim 0.20$, defectors are also driven out from both populations and only cooperators survive, thus a mixture of the cooperator and punisher behaviour, with a frequency of punishment $f_P=p_i\approx 0.41$ in both populations, thus in line with the results in \citep{Xu11} for the one-population SG. On the other hand, as $r$ is raised, a mixture of all three behaviours co-exist, although cooperators take over one population while defectors take over the opponent population. When $r$ is sufficiently high (in the simulation $r>0.956$), defectors take over both populations and cooperation becomes extinct. 

Overall, compared to the one-population SG in \citep{Xu11}, in the two-population SG employing the second mechanism of punishment, although for sufficiently large $r$, cooperation becomes extinct and defectors take over both populations (similar to the effect of spatial structures in one-population SGs as in \citep{Hauert04}), punishment enforced through the cooperation-punishment mixed strategy allows for the extinction of defectors, independently of the initial conditions and the result is valid within the same range of the parameter $r$ as in \citep{Xu11} but with the advantage that stability is always asymptotic. 

\begin{figure}[htp]
\centering
\begin{tabular}{ccc}
\epsfig{file=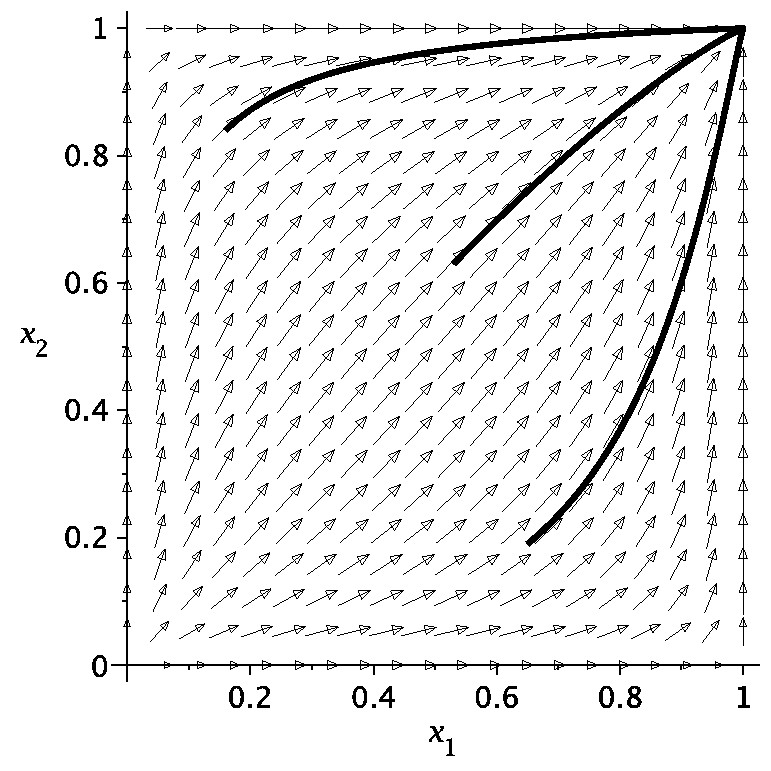,height=5cm,width=5cm,angle=0}&
\epsfig{file=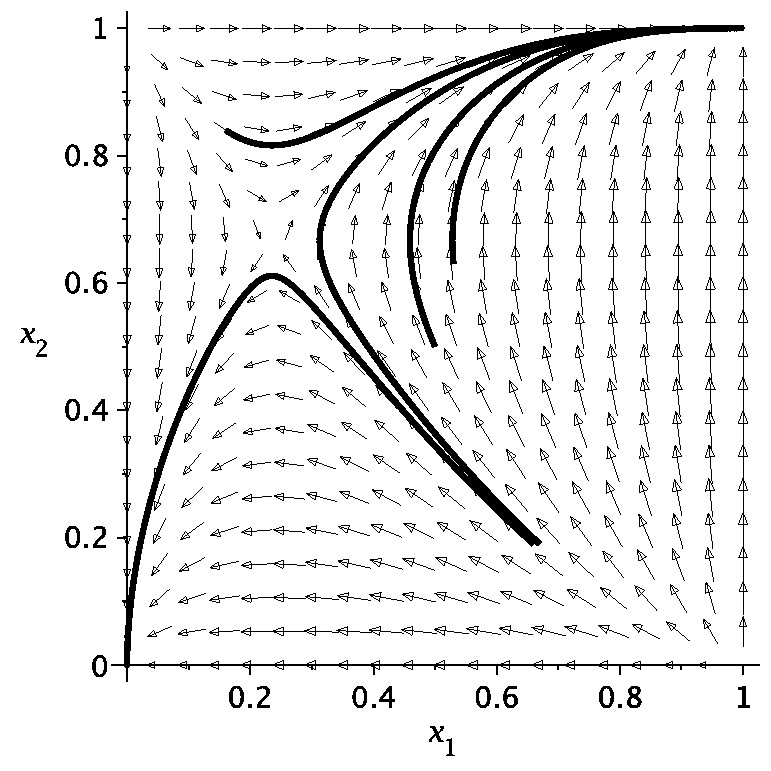,height=5cm,width=5cm,angle=0}&
\epsfig{file=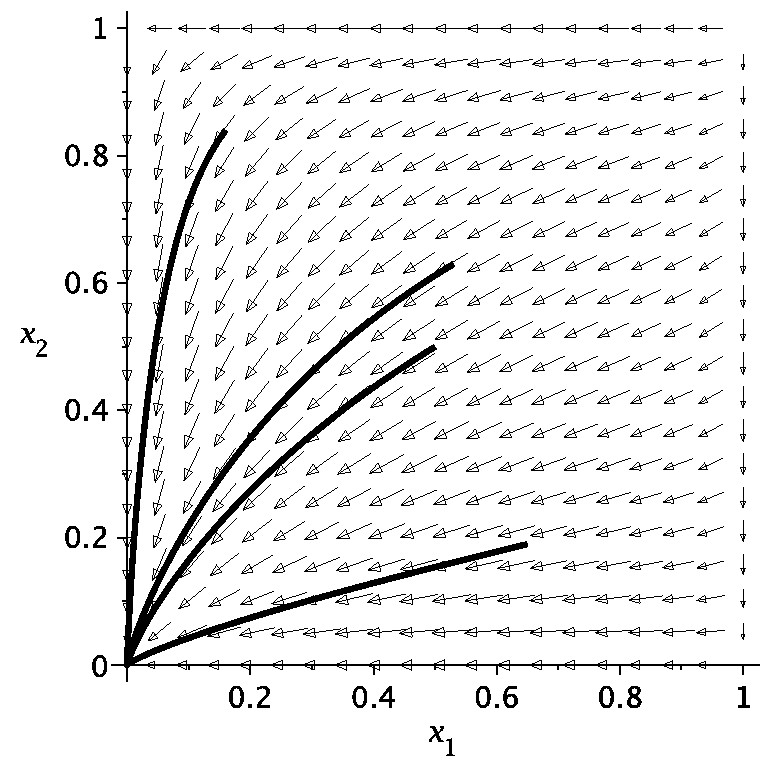,height=5cm,width=5cm,angle=0}\\
\end{tabular}
\vspace{.5cm}
\caption{phase space for $\alpha=0.4$, $\beta=0.9$, $p_1=0.85$, $p_2=0.8$ and $r=0.33$ (left), $r=0.70$ (center) and $r=0.77$ (right).}
\label{fig:3}
\end{figure}

In Figure \ref{fig:3}, I present a simulation in which the likelihood to punish is very high in both populations. I use $\alpha=0.4$, $\beta=0.9$, $p_1=0.85$, $p_2=0.80$ and $r$ varying from $r=0.33$ to $r=0.77$. In this case, the transition between two populations of all-cooperators to two populations of all-defectors has an intermediary situation in which both populations enforce harsh punishment and depending on the initial conditions, all-cooperators in both populations can still be the evolutionary equilibrium even if in the initial state of each population the majority of individuals are defectors.

Figure \ref{fig:4} displays the evolution of cooperators in both populations over time for the case when $r=0.70$ in Figure \ref{fig:3}. Starting with two populations with a share of defectors of 51\% at the initial conditions, cooperators take over both populations in the steady-state. On the other hand, starting with 19\% of cooperators in population 2 and 66\% of cooperators in population 1, defectors take over both populations leading cooperation to extinction. The addition of an extra share of just 1\% of cooperators in population 1 changes the evolutionary pattern completely. In the latter case, the initial conditions cross the separatrix between both basins of attraction of the phase space, leading defectors to extinction instead. For the parameters used in the simulation of Figure \ref{fig:4}, the effect of punishment is to transform the SG into a game of coordination with payoff matrix:
\begin{equation}
\begin{bordermatrix} {&CP&D\cr
		CP & 1,1 &-0.04,0.935\cr
		D & 0.98,-0.02&0,0\cr} 
		\end{bordermatrix}
\label{Matrix-Simul}
\end{equation}
The non-linear system governing the evolution of both populations is composed of equations $\dot{x}_1=x_1(1-x_1)(-0.04+0.06x_2)$ and $\dot{x}_2=x_2(1-x_2)(-0.02+0.085x_1)$. At the initial conditions $(x_1,x_2)=(0.66,0.19)$, $\dot{x}_1<0\wedge\dot{x}_2>0$, thus strategy $CP$ starts to have its share decreased in population 1 while it increases in population 2. In population 1, the share of individuals adopting $CP$ is monotonically decreasing given that $\dot{x}_1$ only changes sign if $x_2>2/3$, which, from Figure \ref{fig:4}, is never achieved. On the other hand, for population 2, one can see from (\ref{Matrix-Simul}) that if the proportion of adopters of strategy $CP$ is sufficiently high in population 1, the best response is to play strategy $CP$, i.e., in the RD natural selection favours individuals who are programmed to behave as cooperators in population 2, thus their share initially increases. Over time, the proportion of adopters of $CP$ keeps falling in population 1 and when the share of the latter falls below $0.235$ (time $t=183$), natural selection starts to favour defectors in population 2. The share of cooperators achieves its peak in population 2 ($x_2=0.6113$) and falls with defectors then increasing monotonically in both populations until the asymptotically stable steady-state $(0,0)$ is reached.

Instead, when the initial conditions are $(x_1,x_2)=(0.67,0.19)$ the dynamics starts in the basin of attraction leading to the steady-state $(1,1)$. In such case, cooperation is monotonically increasing in population 2 while the share of cooperators initially decreases in population 1 until $x_{2}>2/3$ ($t=147$). From that point on, natural selection favours cooperation in both populations and the system flows towards the sink $(1,1)$ with defectors becoming extinct in both populations. The same evolutionary pattern is displayed when initial conditions are $(x_1,x_2)=(0.49,0.49)$, the only difference being that the effect of the recovery of cooperation in population 1 is less pronounced. 

\section{Agent Based Simulation}
\label{sec:ABS}
In this section, I study the second mechanism of punishment using an agent-based simulation (ABS) with $N=60,000$ agents. I extend the methodology used in \citep{Zhong06,Xu11,Wardil10} to two-population games. At the beginning of the simulation, agents are randomly allocated either strategy $CP$ or $D$ and are also randomly allocated to either population 1 or 2, such that the size of each population is $N/2=30,000$ and the shares of cooperators and defectors in each population satisfy some pre-established initial condition. Evolution is modelled as follows: in each time step, two target agents $i_1\in\left\lbrace 1,2,\cdots,N/2 \right\rbrace$ and $i_2\in\left\lbrace 1,2,\cdots,N/2 \right\rbrace$ are randomly selected from populations 1 and 2, respectively. The agents play the game in (\ref{Matrix-III}) against each other and obtain payoffs $V_i^1$ and $V_i^2$, respectively. Meanwhile, two agents $j_1$ and $j_2$ are selected as referencing agents from populations 1 and 2 and the latter also play the game against each other obtaining payoffs $V_j^1$ and $V_j^2$, respectively. The target agents selected from each population $z=\left\lbrace 1,2\right\rbrace $ compare payoffs $V_i^z$ and $V_j^z$. If $V_i^z\geq V_j^z$, the target agent keeps its strategy. If $V_i^z<V_j^z$, the target agent imitates the strategy of the referencing agent with a probability $p_z$ which is proportional to the difference between the payoffs obtained by the target and the referencing agent, $p_z=(V_j^z-V_i^z)/\xi_z$, where $\xi_z$ is the largest entry of the payoff matrix faced by population $z$. Note that $p_z\in\left[0,1 \right] $ given that if the effect of punishment is such that $\nu_z=\min\left\lbrace 1+r-p_{\neg z}\beta,1-r-p_z\alpha,0 \right\rbrace <0$, a positive constant $\gamma=|\nu_z|$ is added to all entries in the payoff matrix of population $z$ such that the ABS is carried out using only positive payoffs, otherwise $\nu_z=0$ and there is no effect on (\ref{Matrix-III}). This transformation does not affect the best-response structure of the game or the steady-states and the RD is not sensitive to it \citep{Weibull97}. After agents $i_z$ have the opportunity of updating their strategies, the time step ends. Updating is asynchronous and when all agents in both populations have on average the opportunity to update their strategies, we call this time span a Monte Carlo time step (MCS). Simulations were carried out with enough MCS to ensure the steady-state was reached. The ABS algorithm was written in FORTRAN programming language and each simulation was repeated 100 times with the same number of MCS to take into account random effects.   

\begin{figure}[htp]
\centering
\begin{tabular}{ccc}
\epsfig{file=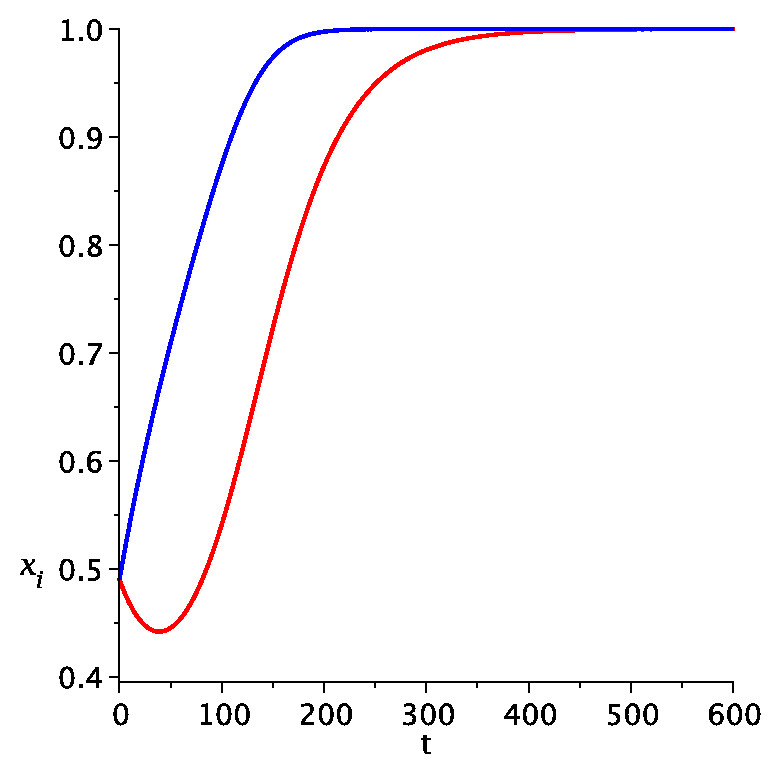,height=4.75cm,width=5cm,angle=0}&
\epsfig{file=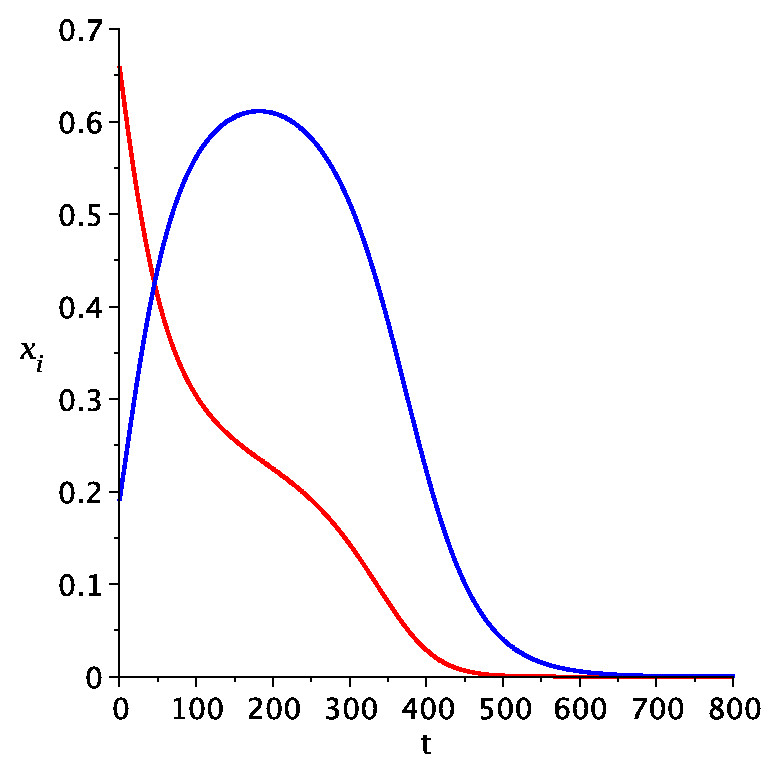,height=4.75cm,width=5cm,angle=0}&
\epsfig{file=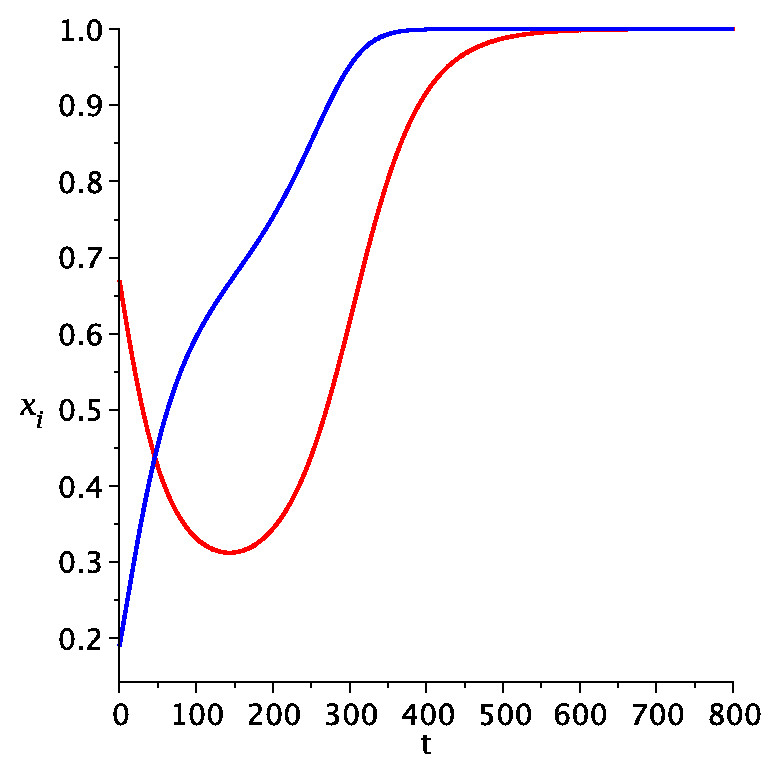,height=4.75cm,width=5cm,angle=0}\\
\end{tabular}
\vspace{.5cm}
\caption{(color online) evolution of individuals adopting strategy $CP$ in populations 1 (red) and 2 (blue) over time; from left to right, $\alpha=0.4$, $\beta=0.9$, $p_1=0.85$, $p_2=0.80$ , $r=0.70$ and initial conditions $(0.49,0.49)$ (left), $(0.66,0.19)$ (center) and $(0.67,0.19)$ (right).}
\label{fig:4}
\end{figure}

\begin{figure}[htp]
\centering
\begin{tabular}{cc}
\epsfig{file=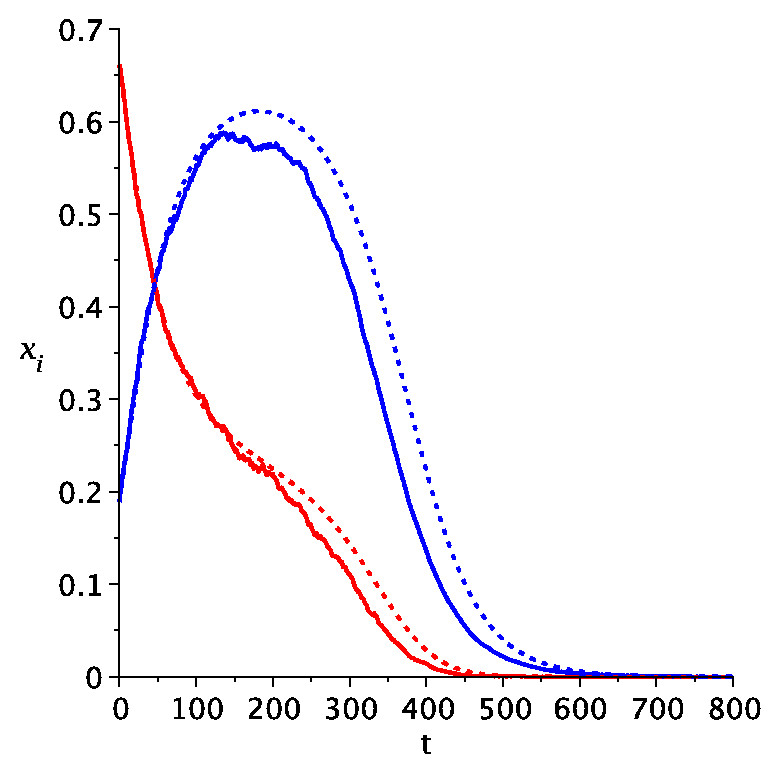,height=4.75cm,width=6cm,angle=0}&
\epsfig{file=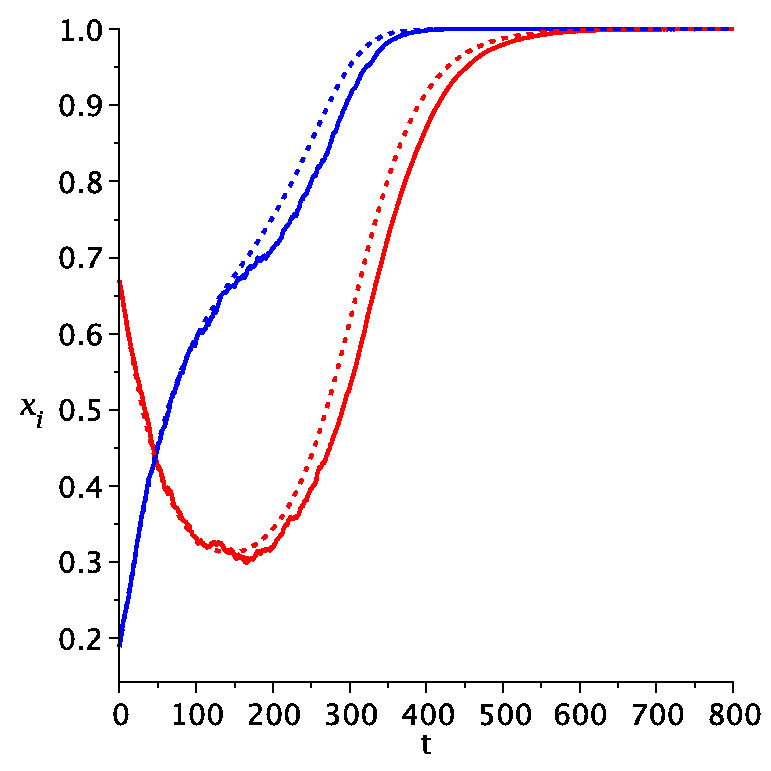,height=4.75cm,width=6cm,angle=0}\\
\end{tabular}
\vspace{.5cm}
\caption{(color online) two typical runs for the evolution of strategy $CP$ in populations 1 (red) and 2 (blue) over time based both on ABS (solid lines) and on differential equations (dot lines), $\alpha=0.4$, $\beta=0.9$, $p_1=0.85$, $p_2=0.80$ , $r=0.70$ and initial conditions $(0.665,0.19)$.}
\label{fig:5}
\end{figure}

In Figure \ref{fig:5}, I present the typical runs of the ABS code for the two cases presented in Figure \ref{fig:4} with initial conditions $(0.66,0.19)$ and $(0.67,0.19)$. The ABS used the same parameters but with initial conditions $(0.665,0.19)$, i.e., initial conditions located ``on the separatrix'' between both basins of attraction. Due to the probabilistic nature of the simulation, 61 repetitions led to the extinction of defectors in both populations while the remaining 39 repetitions led to the extinction of cooperators. The diagrams in Figure \ref{fig:5} show a good level of agreement between the ABS (solid lines) and the dynamics based on the non-linear system of ODEs (dot lines), regarding both the evolutionary pattern over time and the steady-states.    

\begin{figure}[H]
\centering
\begin{tabular}{cc}
\epsfig{file=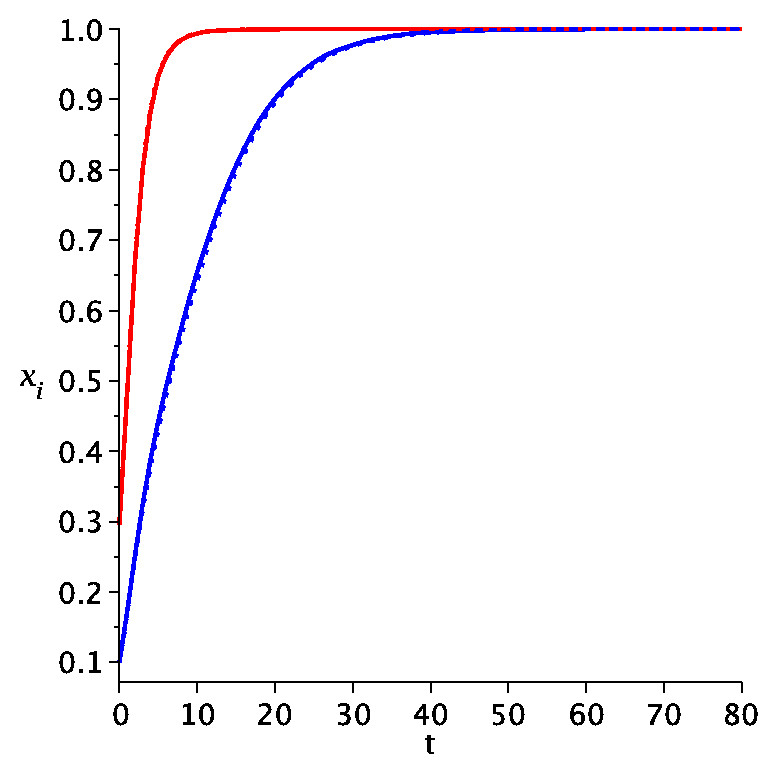,height=5cm,width=6cm,angle=0}&
\epsfig{file=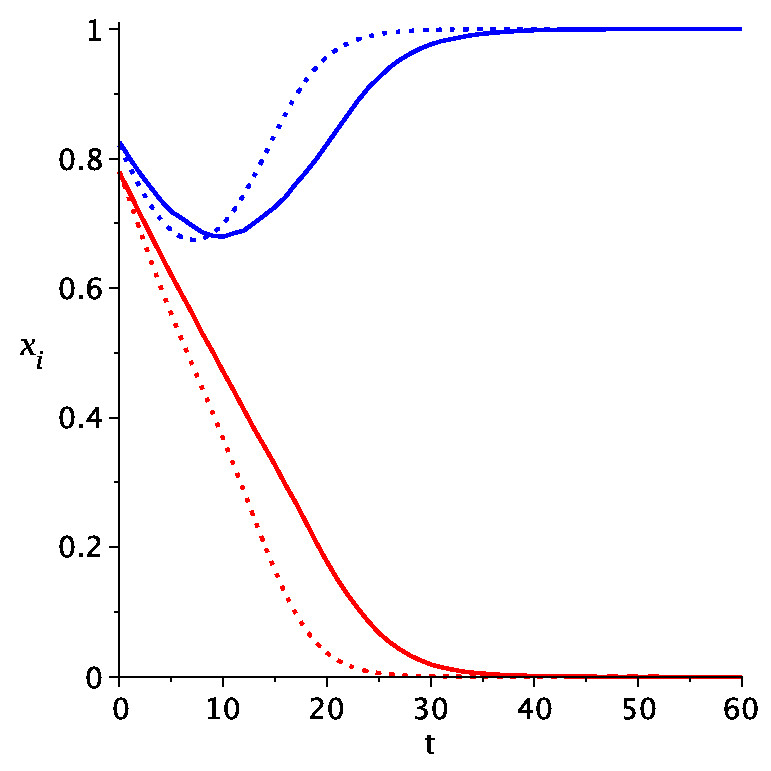,height=5cm,width=6cm,angle=0}\\
\end{tabular}
\vspace{.5cm}
\caption{(color online) typical runs for the evolution of strategy $CP$ in populations 1 (red) and 2 (blue) over time based both on ABS (solid lines) and on differential equations (dot lines), $\alpha=0.1$, $\beta=0.5$, $p_1=0.41$, $p_2=0.405$. Left panel: $r=0.05$ and initial conditions $(0.3,0.1)$. Right panel: $r=0.60$ and initial conditions $(0.78,0.82)$.}
\label{fig:6}
\end{figure}

Two additional ABS were carried out, both using the same parameters of cases presented in Figure \ref{fig:2}, i.e., $\alpha=0.1$, $\beta=0.5$, $p_1=0.41$ and $p_2=0.405$. The first simulation used $r=0.05$ and initial conditions $(0.3,0.1)$ while the second simulation used $r=0.60$ and initial conditions $(0.78,0.82)$. Again, in both cases there was good agreement between the ABS and the results that were obtained analytically using (\ref{RD-II-x}) and (\ref{RD-II-y}) with regard to the steady state and the evolutionary pattern over time. In the case of the simulation with $r=0.60$, all 100 repetitions led to the same steady-state given the initial conditions are located far enough from the separatrix between both basins of attraction of the phase space. Figure \ref{fig:6} presents typical runs of these two simulations. 
\section{Conclusion}
\label{sec:conclusion}
The emergence of cooperation was studied in a two-population snowdrift game using two different mechanisms based on punishment. The first mechanism followed the traditional approach in the literature in which a third strategy is added to the game. In this case, I extended the framework in \citep{Xu11} to the case of two well-mixed populations contesting against each other. A non-standard mechanism was then considered, in which the strategy to cooperate was modified in order to allow cooperators to be able to punish defectors with a given exogenous frequency. The latter was equivalent to replace pure strategy $C$ by a mixed strategy in which players randomize between cooperation and punishment. In both cases, defectors became extinct if the probability of punishment was above $r/\beta$. On the other hand, although defectors could become extinct in both populations under the two different mechanisms of punishment, the second mechanism led to a more robust equilibrium given that the latter was asymptotic instead of the neutrally stable equilibrium attained by the first mechanism of punishment. Moreover, from a purely mathematical point of view, the second mechanism had the advantage that the dynamics of the game could be studied in the $\Re^2$-space instead of the less-intuitive $\Re^4$-space. Evolutionarily speaking, in the second mechanism, whenever cooperators were to take over both populations, extinction of defectors was a globally asymptotically stable state for any interior initial conditions in the phase space. Results were obtained analytically using a non-linear system of ordinary differential equations and were then compared with results from an agent-based simulation, which displayed a good level of agreement with the former. Regarding future research, two immediate extensions would be to consider a spatial game in a squared lattice with Moore neighbourhood and periodic boundary conditions or to take into account endogenous (frequency-dependent) payoffs in the stage-game.

\section*{Acknowledgements}
I acknowledge the research support ``Programa de Incentivo \`a Produtividade em Ensino
e Pesquisa'' from PUC-Rio.


\end{document}